# Design, implementation, and on-sky performance of an advanced apochromatic triplet atmospheric dispersion corrector for the Magellan adaptive optics system and VisAO camera

Derek Kopon[1,2], Laird M. Close[2], Jared R. Males[2], and Victor Gasho[2]

[1]Max Planck Institute for Astronomy, Koenigstuhl 17, 69117 Heidelberg, Germany; derek.kopon@gmail.com

[2]Center for Astronomical Adaptive Optics, Steward Observatory, University of Arizona, Tucson, AZ USA 85721

## ABSTRACT

We present the novel design, laboratory verification, and on-sky performance of our advanced triplet atmospheric dispersion corrector (ADC), an important component of the Magellan Adaptive Optics system (MagAO), which recently achieved first light in December 2012. High-precision broadband ($0.5 - 1.0$ μm) atmospheric dispersion correction at visible wavelengths is essential both for wavefront sensing (WFS) on fainter guide stars, and for performing visible AO science using our VisAO science camera. At 2 airmasses (60° from zenith) and over the waveband 500-1000 nm, our triplet design produces a 57% improvement in geometric rms spot size, a 33% improvement in encircled energy at 20 arcsec radius, and a 62% improvement in Strehl ratio when compared to a conventional doublet design. This triplet design has been fabricated, tested in the lab, and integrated into the MagAO WFS and the VisAO science camera. We present on-sky results of the ADC in operation with the MagAO system. We also present a zero-beam-deviation triplet ADC design, which will be important to future AO systems that require precise alignment of the optical axis over a large range of airmasses in addition to diffraction-limited broadband dispersion correction.

## 1. Introduction: The Need for an Advanced ADC

Current adaptive optics (AO) imaging systems on large ground-based telescopes are just now approaching truly diffraction-limited high-Strehl performance in the visible and near-IR bands (Close et al. 2010 and 2013). Consequently, heretofore-ignored effects, such as secondary color caused by atmospheric dispersion, will need to be mitigated very precisely. Many of the science cases for ground-based diffraction-limited imaging are photon-starved (such as the imaging of faint dust disks or the detection of faint extrasolar planets), thereby driving the need for broadband imaging. To achieve these science goals, current 8-meter class and future 30-meter class telescopes will require well-corrected broadband chromatic performance (Goncharov et al. 2007 and Devaney et al. 2008). Additionally, increased sky coverage can be achieved if AO can be performed on fainter guide stars, thereby also driving the need for atmospheric dispersion correction over as broad a band as possible (0.5 – 1.0 μm) to maximize the light on the WFS CCD.

The Magellan Clay Telescope is a 6.5-meter Gregorian telescope located in Las Campanas Observatory (LCO) in Chile. The Magellan AO system consists of an 850 mm diameter thin shell (1.6 mm thick) adaptive secondary mirror (ASM) with 585 actuators, a pyramid wavefront sensor (PWFS), a 1-5 μm infrared science camera called Clio2 (Sivanandam et al. 2006), and our VisAO visible (500-1000 nm) science CCD camera (Kopon et al. 2009; Males et al. 2012). The secondary mirror and wavefront sensor are the same in design and optical prescription as those of the LBT (Esposito et al. 2010). A dichroic beamsplitter (the tilted Clio2 window) located before the Nasmyth focus allows us to simultaneously perform IR and visible science. The ADC provides common path chromatic correction on the visible light beam for both the pyramid wavefront sensor and the VisAO science camera. For more information on the Magellan AO system see Close et al. (2012).

The PWFS and VisAO camera are both mounted on the same optical board called the W-unit (Figure 1), which receives visible light from the dichroic (Kopon et al. 2009 and 2010). The W-unit is mounted on a stable and precise 3-axis stage that patrols a 2.3x3.2 arcmin field of view on the sky in order to acquire guide stars. After the f/16 Magellan Clay focal plane, light from the telescope passes through a triplet lens that changes the beam from diverging f/16 to converging f/52. We call this apochromatic triplet lens the "input lens." It is similar to that used on the LBT W-Units (Esposito et al. 2010). This beam then passes through the ADC before hitting a beamsplitter/dichroic wheel. The transmitted light from the beamsplitter goes to the PWFS and the reflected light goes to the VisAO camera. Figure 2 shows the raytrace for the VisAO science arm and the PWFS arm of the W-unit.

The PWFS is a new method of wavefront sensing (Esposito et al. 2000) that has significant advantages over conventional methods such as the Shack-Hartmann sensor or the curvature sensor (Roddier et al. 1988). The PWFS operates by using a fast piezo scan mirror to circularly modulate an image of the guide star around the tip of a 4-sided optical pyramid that is located at the f/52 focal plane. Light passes through each side of the pyramid and forms four pupil images in the quadrants of the PWFS CCD detector. By changing the radius of modulation down to ±2 λ/D (for bright stars) and/or binning the CCD from 28x28 pix/pupil to 4x4 pix/pupil for faint stars, the dynamic range of the PWFS can be easily adjusted based on the science case and guide star magnitude, a significant advantage over the Shack-Hartmann method, which has a spatial sampling fixed by the lenslet pitch. To take full advantage of the dynamic range of the pyramid sensor, a modulation radius approaching the FWHM of the diffraction-limited star image (~30 mas at D = 6.5 m) is necessary for very fine wavefront sensing. This requires a high-performance ADC, which enables finer broadband wavefront sensing on fainter guide stars than would otherwise have been possible at moderate airmass.

The atmospheric dispersion over the band 0.5-1.0 μm results in a PSF that is diffraction-limited (39-78 mas diameter between the first Airy minima) in the direction parallel to the horizon but elongated by ~1,700 mas in the perpendicular direction (Figure 3) at 60° from zenith (2 airmasses)! The mitigation of this dispersion is essential for both wavefront sensing and diffraction-limited broadband visible AO science (even at airmasses much lower than 2). Traditional linear or 2-doublet ADC designs (also known as Amici prisms) can significantly reduce this 1.7 arcsec of lateral color by correcting primary color (see for example Phillips et al. 2006 or Wynne 1997), but still do not reach the diffraction limit because of secondary and higher order lateral color (Wallner et al. 1980).

## 2. The triplet design

Most ADCs designed and built to date consist of two identical counter-rotating prism doublets made of a crown and flint glass (for example see Wynne 1997), also known as Amici prisms, or zero-deviation prisms. One method of designing a conventional 2-doublet ADC in the visible band is to match the d-line (587 nm) indices of the two glasses as closely as possible in order to avoid steering the beam away from its incident direction while allowing the faces of the ADC to remain parallel. The wedge angles and glasses of the prisms are chosen to correct primary chromatic aberration at the most extreme zenith angle. By then rotating the two doublets relative to each other, an arbitrary amount of first-order chromatic aberration can be added to the beam to exactly cancel the dispersive effects of the atmosphere at a given zenith angle. The 2-Doublet design corrects the atmospheric dispersion so that the longest and shortest wavelengths overlap each other, thereby correcting the primary chromatism. Secondary chromatism is not corrected and is the dominant source of error at higher zenith angles.

The 2-Doublet design that we compare our performance to is that initially designed by the Arcetri group for use with the LBT AO system. The prescription for this design is presented in Table 1. The glasses used in this design, Corning's C04-64 and Ohara's BAM23 (at present time these have become obsolete glasses and are no longer in production), are very well matched, with d-line indices of 1.6035 and 1.6072, respectively. The Abbe numbers are well matched to be symmetrically on either side of 50: $V_d$ = 63.402 for C04-64 and $V_d$ = 40.263 for BAM23. This doublet design corrects first-order chromatism over the design wavelengths very well. To correct higher orders of chromatism, more glasses and thereby more degrees of freedom are needed.

**Table 1**

**Doublet ADC Optical Prescription**

| Surface | Thickness (mm) | Glass | Tilt Angle (deg) |
|---------|----------------|-------|------------------|
| 1 | 4 | C04-64 (Corning) | 0 |
| 2 | 4 | BAM23 (OHARA) | 28.81 |
| 3 | … | … | 0 |

To optimize the glass choice for our triplet design, we performed a Zemax glass substitution optimization that entailed swapping glasses in and out and then optimizing the wedge angles to minimize rms spot size at the maximum design zenith angle. Some designs were then discarded because the beam deviation was too large, or the required wedge angles were too large to realistically fabricate without very thick ADCs. The three glasses we ultimately chose for our design are Ohara's S-PHM53 (crown), S-TIM8 (flint), and Schott's N-KZFS4 (anomalous dispersion). These glasses have well matched indices, with $n_d$ = 1.603, 1.596, and 1.613, respectively. The Abbe numbers are also well matched, with V = 65.44, 39.24, and 44.49, respectively. The N-KZFS4 has one of the highest anomalous dispersion figures of merit in the glass catalogue. The figure of merit used in Zemax is the $DP_{g,F}$ value, which measures the deviation of the partial dispersion versus Abbe number from the "normal" line, which fits most glasses. N-KZFS4 has $DP_{g,F}$ = -0.01, compared to 0.0045 for S-PHM53 and 0.0023 for S-TIM8, respectively. Table 2 contains the optical prescription of our triplet ADC. Figure 4 shows drawings of the 2-doublet and 2-triplet designs. For more on anomalous dispersion definitions and glass choice, see http://www.radiantzemax.com/kb-en/KnowledgebaseArticle50231.aspx.

**Table 2**

**Triplet ADC Optical Prescription**

| Surface | Thickness (mm) | Glass | Tilt Angle (deg) |
|---------|----------------|-------|------------------|
| 1 | 5 | S-PHM53 (OHARA) | 0 |
| 2 | 3 | S-TIM8 (OHARA) | 32.21 |
| 3 | 4 | N-KZFS4 (SCHOTT) | 24.51 |
| 4 | … | … | 0 |

The Zemax optical design software has the ability to simulate the chromatic effects of the atmosphere by placing a model optical surface in front of the Magellan Clay telescope design. This surface simulates atmospheric dispersion with adjustable input parameters for Zenith angle, temperature, elevation, atmospheric pressure, humidity, etc. In our design process, the atmospheric surface was set to 70 degrees zenith and the relative angles of the ADC were set to 180 deg. The wedge angles of the three prisms in the triplet were then optimized to correct both primary and secondary color. More details on this design, including a chromatic pupil shear analysis that quantifies beam displacement of the chief ray are presented in Kopon et al. 2008.

## 3. Design comparison

Figure 5 shows the geometric spot diagrams for the two designs over the band 0.5-1.0 μm at 60° from zenith. In the spot of the doublet design, the longest and shortest wavelengths overlap, indicating that primary chromatism is well corrected and that the dominant aberration is secondary chromatism. Because the magnitude of the chromatic aberration in the geometric spot diagrams of the two designs is on the same order of magnitude as the diffraction limit of the telescope, we present the broadband diffraction PSFs of the two designs in Figure 6. These PSFs are computed using the Fast Fourier Transform in a Zemax model that includes the full telescope and all optics, the central obscuration from the secondary, and the spider vanes. Note the superior diffraction-limited PSF provided by the triplet design. The improvement resulting from this design will be quantified next.

Figures 7-9 compare the performance of the two designs as a function of zenith angle. Figure 7 shows polychromatic Strehl ratio. Figure 8 shows encircled energy inside of the radius of the theoretical first Airy minimum at λ = 500 nm. Figure 9 shows the FWHM of the broadband diffraction FFT PSF in the direction of the atmospheric dispersion (perpendicular to the horizon).

## 4. Laboratory performance

When used on-sky looking at a broadband point source, such as a star, the ADC will be taking a spectrum of atmospherically dispersed light from the Magellan telescope f/16 focal plane and correcting it so that it falls on the CCD47 at the VisAO f/52 focal plane and the pyramid tip of the PWFS as a well-corrected broadband point. However, in our lab we cannot easily and reliably simulate the dispersive effects of the atmosphere in order to generate that low resolution spectrum. Therefore, we have designed a test that works in reverse: the ADC takes a white-light point source from a pinhole located where the telescope f/16 focal plane would be and disperses it into a spectrum at the lab detector f/52 focal plane (Kopon et al. 2012). Using narrow band filters, we measure where different wavelengths fall on the focal plane and compare these displacements to the Zemax predictions. The three filters we chose were off-the-shelf laser line filters from Thorlabs with central wavelengths of 532 ± 2 nm (10 ± 2 nm FWHM), 850 ± 5 nm (10 ± 2 nm FWHM), and 905 ± 5 nm (25 ± 5 nm FWHM).

Using a microscope objective and a pair of achromatic doublets, we reimage a fiber white-light source onto a 10μm pinhole. This pinhole serves as the point source that is located where the nominal Magellan telescope f/16 focal plane would be. The point source feeds the W-unit triplet input lens, which converts the beam from a diverging f/16 beam to a converging f/52 beam. This converging white light beam then passes through the ADC triplet prisms, which are in rotating mounts. The image is then measured with our Electrim EDC-3000D lab CCD. By placing three different narrow band filters in front of the white light source (532 nm, 850 nm, and 905 nm) and measuring where these wavelengths fall on the focal plane relative to each other for various relative ADC clockings, we are able to measure both the primary and secondary dispersion characteristics of the ADC (Figure 10).

The comparison of this test and the Zemax theoretical predicted curves are shown in Figure 11. We expect that the most sensitive results are given by the largest ADC relative clocking angle and the largest wavelength difference. In these cases, the measured displacement of the 532 nm and 905 nm fiber images at the lab CCD focal plane differs from the Zemax prediction by only 0.45± 0.15%, which is on the order of the estimated error of our lab setup. Therefore, the ADC dispersion is correct to within our measurement error, confirming that the ADCs were fabricated correctly, are chromatically well behaved, and perform as expected.

## 5. Magellan AO on-sky performance

The Magellan AO system achieved first light in December of 2012 at LCO in Chile. For details of the MagAO commissioning and performance, see Close et al. 2013. On the night of December 3[rd] (UT), we closed the AO loop on Theta 1 Orionis B with the AO system correcting 200 modes and operating at 1 kHz. Seeing was 0.7 arcsec. The source was located 30° from zenith (1.15 airmasses). Figure 12 shows a 30 second exposure in i'-band (710-840 nm) on Theta Ori B with both the ADC "off" (i.e. in its zero-dispersion configuration with both clocking angles set to zero), and "on" (rotated to the optimal angle to correct atmospheric dispersion – see Appendix 1). The ADC "on" image has a Strehl ratio of 7% and a FWHM of 32 mas compared to the Zemax predicted perfect FWHM of 24.3 mas in the i'-band.

It is clear from these images that an optimally tuned ADC is required for diffraction-limited imaging in the visible even at relatively small airmasses of 1.15. It is even more important to correct atmospheric dispersion for the PWFS which is operating "wide open" (600-1000 nm). The MagAO PWFS and the VisAO camera share the same ADC, eliminating the need for 2 separate ADCs.

Appendix 1 contains a lookup table of the triplet ADC clocking angles specific to typical Magellan site parameters. This table is now used continuously with nightly operations of the AO system.

The triplet ADC has proven to be an indispensible optical component for the MagAO system. In particular, as is clearly shown by Fig. 12, high resolution broad band VisAO science would be impossible without it. This paper offers a direct demonstration that any future visible light (600-1000 nm) AO system will need an accurate ADC to work at or near the diffraction limit of a large telescope.

## 6. The zero-deviation design

An ADC can be designed to put zero angular deviation to the chief ray at all angles of rotation by allowing the faces of the ADC to have some wedge (Wynne 1997). In the triplet design that we fabricated for the Magellan AO system, we made the faces parallel for ease of manufacture and chose the glasses of the triplet in order to meet our dispersion requirements and minimize angular beam deviation. However, for any combination of glasses, there will still be some non-zero amount of angular beam deviation. Our triplet design has a factor of 10 lower angular deviation (1.96 arcmin mechanical deviation as the beam passes through the ADC at 60° from Zenith, which is 149 mas on the sky) compared to the doublet design (18.8 arcmin mechanical beam deviation, which is 1.31 arcsec on the sky). In any case this small (<0.2 arcsec of tilt) motion doesn't effect the VisAO camera since it is sensed and removed at 1KHz by the PWFS which is common path with the VisAO camera (Fig 1). It does however, slightly changes the position of the star on Clio2 (which is upstream of the ADC, and so is non-common path w.r.t. the ADC tilt.). However, this is a very small tilt that takes typically >1 hour to build up (as tracking changes Zenith angle by 60 deg). Hence, we know this slow <2 mas/min motion doesn't effect a typical single Clio2 exposure which is always less than 30 seconds. So at most a single Clio2 image could be degraded ~1 mas which insignificant compared to the PSF FWHM > 40 mas for Clio2 from 1.2-5.3 microns.

Yet even though this beam deviation is a non-issue for MagAO, many other applications could require an optical design with no angular beam deviation. For instance, our pyramid wavefront sensor has a pupil rerotator that is precisely aligned to the optical axis of the system, with a very tight tolerance on the angle at which it can accept the incoming beam. An ADC with significant beam deviation can throw the rerotator out of alignment and prevent the wavefront sensor from working at higher airmass. Another example of beam deviation becoming a problem would be a coronagraph or a spectrograph with a narrow input slit that requires precise alignment of the star to the coronagraphic mask or slit during long integrations.

Therefore, in Table 3 we present a "zero-deviation" ADC design with wedged faces that uses the same glass as our original triplet design and is optimized to have zero beam deviation at all zenith angles up the 70° from zenith design limit. Figure 13 shows beam deviation at the f/52 focal plane as a function of zenith angle for the doublet, triplet, and zero-deviation triplet design.

**Table 3**

**Triplet Zero-Deviation ADC Optical Prescription**

| Surface | Thickness (mm) | Glass | Tilt Angle (deg) |
|---------|----------------|-------|------------------|
| 1 | 5 | S-PHM53 (OHARA) | 0.186 |
| 2 | 3 | S-TIM8 (OHARA) | 32.497 |
| 3 | 4 | N-KZFS4 (SCHOTT) | 24.844 |
| 4 | … | … | 0.231 |

## 7. Conclusions

We discuss the motivation, design, fabrication, laboratory verification, and on-sky performance of our advanced apochromatic triplet ADC for the Magellan AO system. The ADC performs in the lab as we predicted and this performance was verified on-sky during the December 2012 commissioning of the MagAO system. We also present the design of our zero-deviation triplet ADC, which gives the same chromatic performance as our original triplet design and zero beam-deviation at all design zenith angles. We expect that the design principles and techniques that we have employed in this project will be useful for future high-resolution ground-based systems.

## Acknowledgements

We would like to thank the NSF MRI, TSIP, and ATI programs for their generous support of this project. We would also like to thank our partners and collaborators at the Magellan Observatory, the Carnegie Institute, and Arcetri Observatory. We thank Optimax, who fabricated the triplet ADCs and were up for the challenge of meeting our unique and exacting design specifications.

## References and links

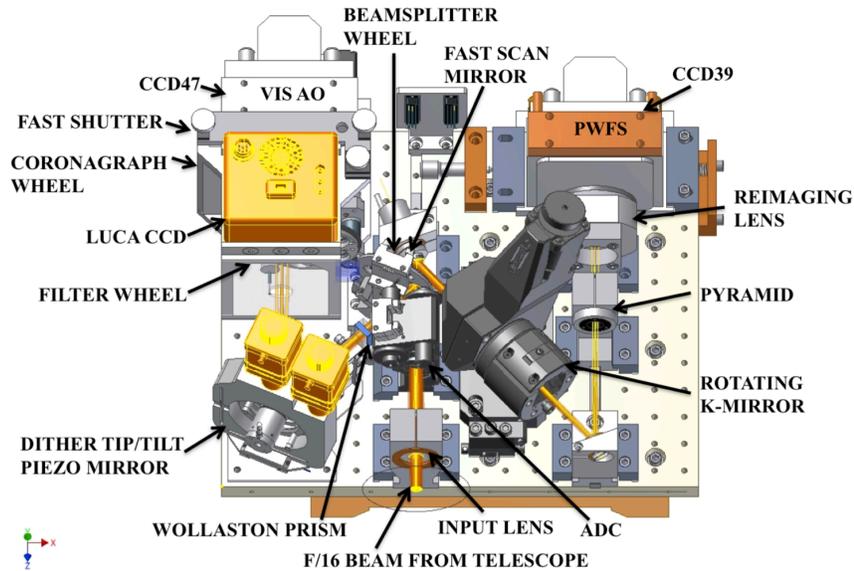

**Fig. 1.** The W-unit optical board. Incoming light from the telescope (center bottom) passes through the ADC before hitting the beamsplitter wheel. Reflected light from the beamsplitter goes to the VisAO science camera. Transmitted light goes to the PWFS. The ADC is common-path for both the PWFS and the VisAO science camera.

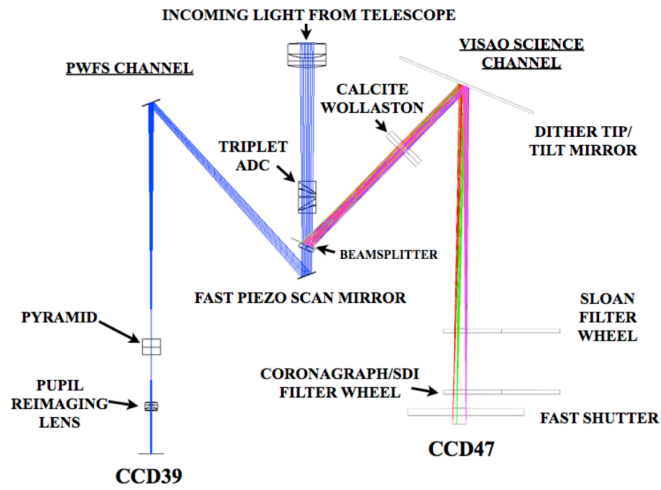

**Fig. 2.** Raytrace of the W-unit optical board containing the VisAO science camera (right hand side) and PWFS (left hand side) optical channels.

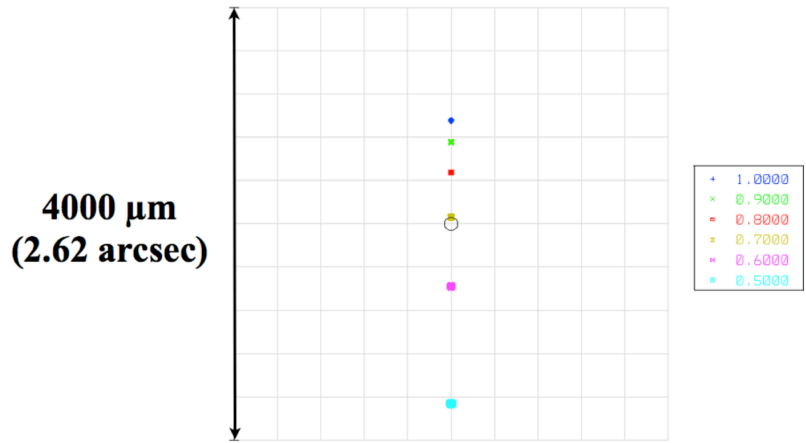

**Fig. 3.** Uncorrected atmospheric dispersion at the VisAO f/49 focal plane at 60° from Zenith (2 airmasses) over the band 0.5-1.0 μm. The small circle in the center represents the first Airy minimum of the diffraction limited PSF at λ = 1 μm. Units are in microns with a plate scale of 0.595 mas/μm at the VisAO f/49 focal plane.

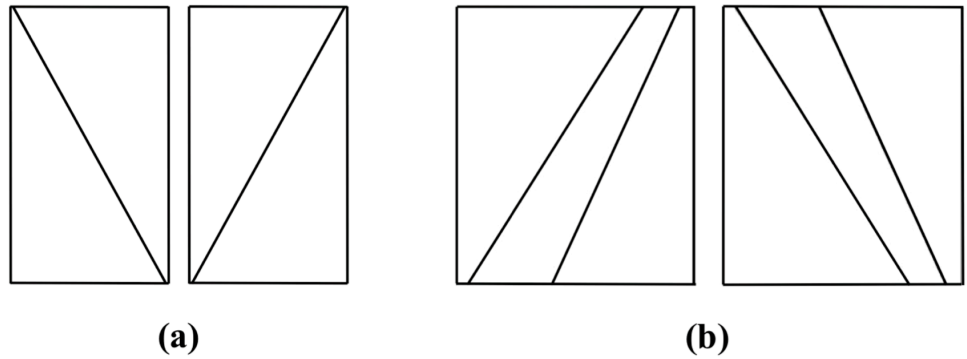

**Fig. 4.** (a) The conventional 2-doublet "Amici Prism" ADC. (b) Our 2-triplet design. Light enters from the left. The ADCs are shown rotated to be in their "zero-dispersion" configurations. The optical prescriptions for these designs are given in Tables 1 and 2.

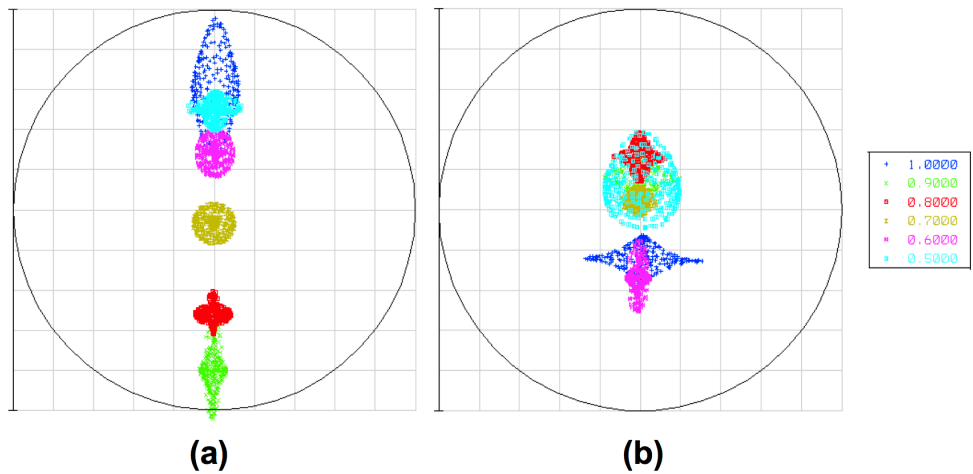

**Fig. 5.** (a) Geometric spot diagram of the 2-doublet design over the band 0.5-1.0 microns at 60° from zenith. (b) Geometric spot diagram of our 2-triplet design at the same wavelengths and zenith angle giving a 57% improvement in rms spot size. The circle shows the first airy minimum in the diffraction limit at λ = 1.0 μm.

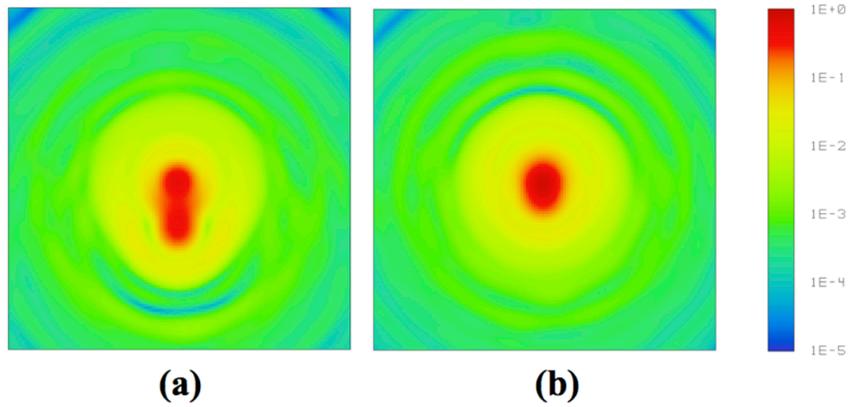

**(a)**                    **(b)**

**Fig. 6.** (a) Broadband (0.5-1.0 μm) PSF for the doublet design computed with Zemax using the Fast Fourier Transform (FFT) method to account for diffraction effects. (b) FFT PSF for the triplet design.

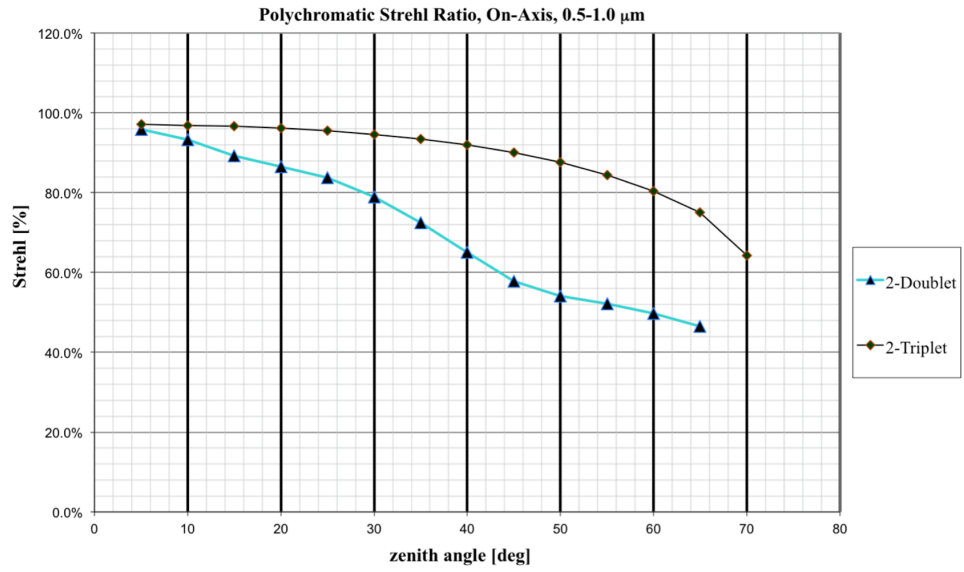

**Fig. 7.** Polychromatic Strehl ratio for the two designs as a function of zenith angle. Perfect AO correction is assumed here for these comparisons.

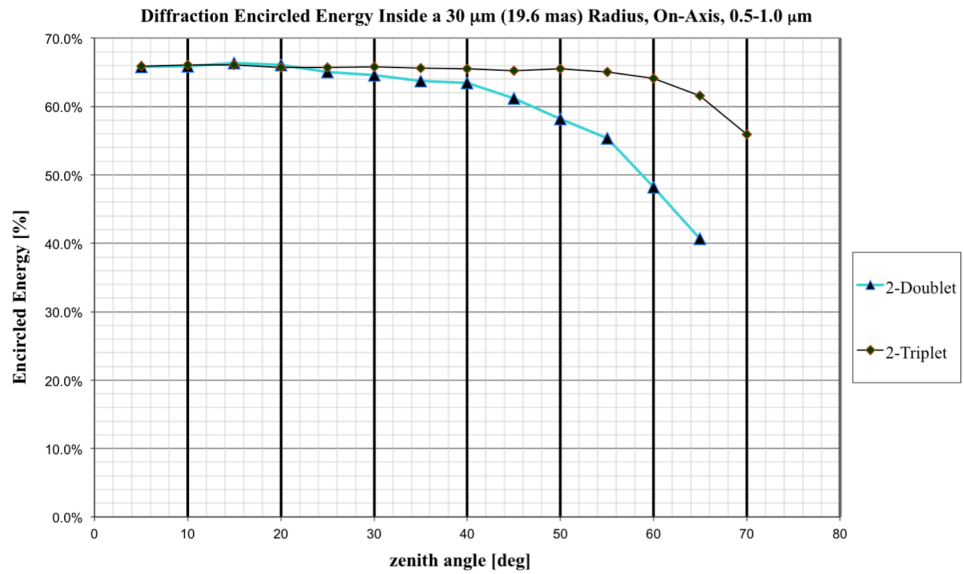

**Fig. 8.** Diffraction encircled energy as a function of zenith angle for the two designs.

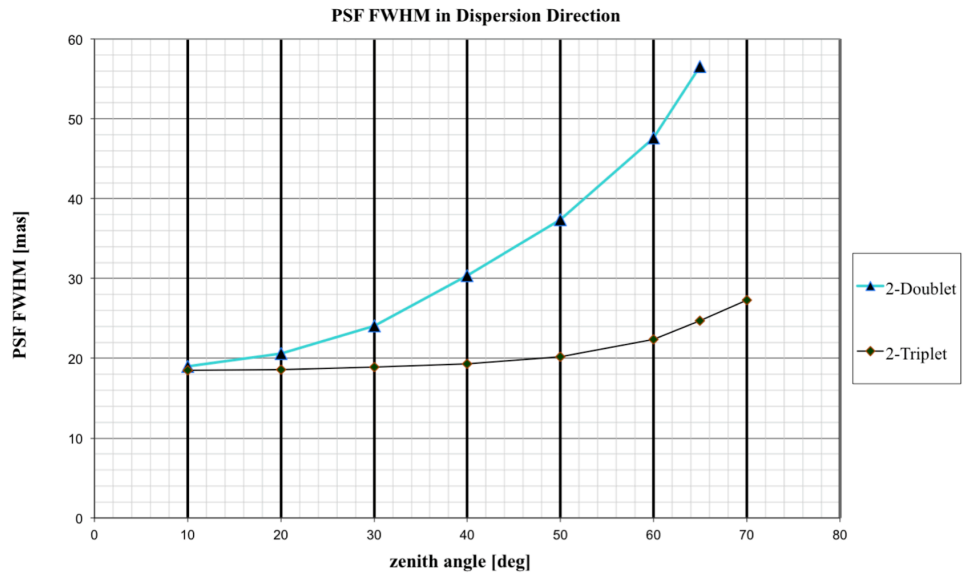

**Fig. 9.** FFT PSF FWHM in the dispersion direction as a function of zenith angle for the two designs.

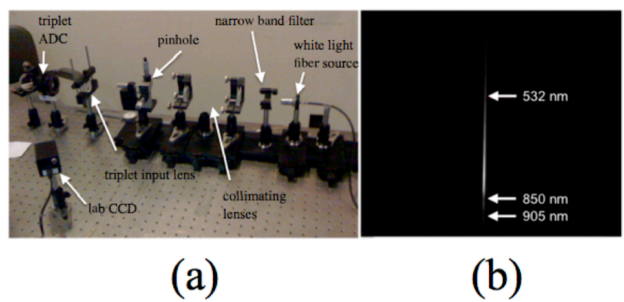

**Fig. 10.** (a) Our lab setup for testing the dispersion characteristics of our triplet ADC. A white light fiber source is collimated and reimaged on a pinhole, which creates an F/16 beam to feed our input lens. The ADCs are rotated to their maximum dispersion position and narrow band filters are used to select certain wavelengths. (b) The white light source dispersed into a spectrum by the ADCs in their maximum dispersion position. The locations of three wavelengths were measured and compared to the Zemax prediction.

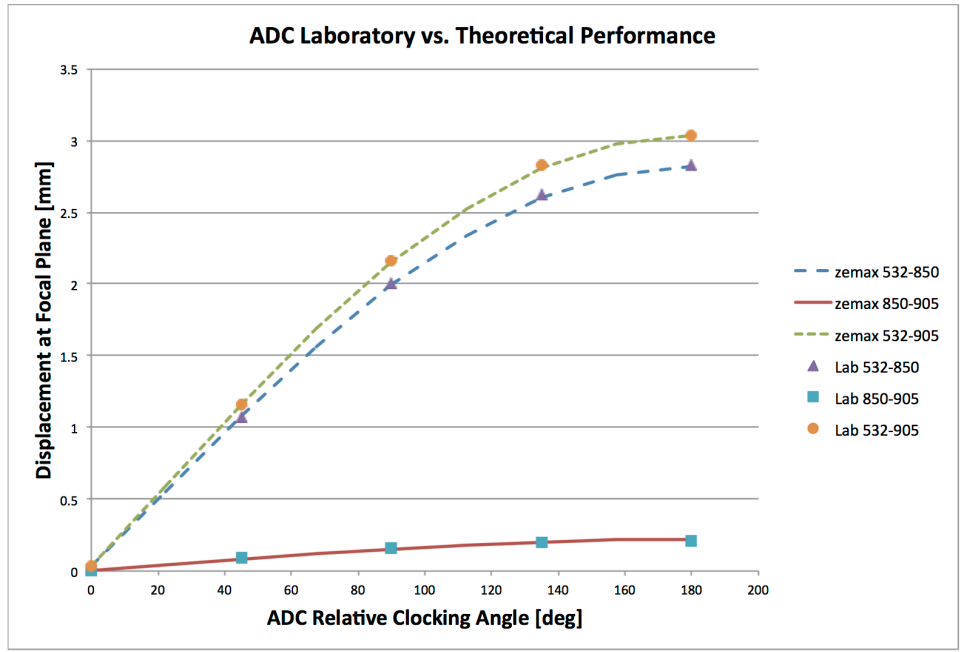

**Fig. 11.** Theoretical vs. measured displacement at the focal plane of three different wavelengths dispersed by the ADC in the lab setup shown in Figure 10.

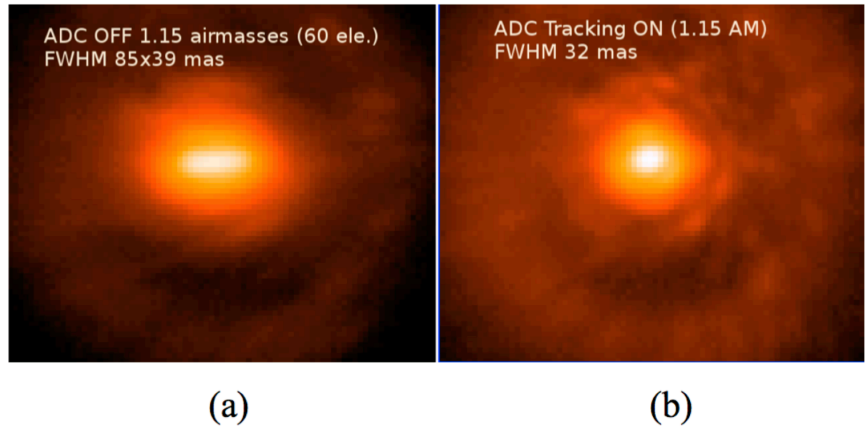

(a) (b)

**Fig. 12.** (a) MagAO corrected VisAO i' (765 nm) 60s image of Theta 1 Orionis B with the ADC in its zero-dispersion "OFF" configuration during 0.7" seeing at 1.15 airmass. (b) Another 60s image of the same star taken immediately after (also at 1.15 airmasses) but with the ADC correcting atmospheric dispersion (ADC tracking "ON"). Note that even a relatively narrow filter like i' (710-840 nm) has significant dispersion at just 1.15 airmasses once visible AO is used on large telescopes. Log stretch, 0.3x0.3" FOV, elevation axis is horizontal.

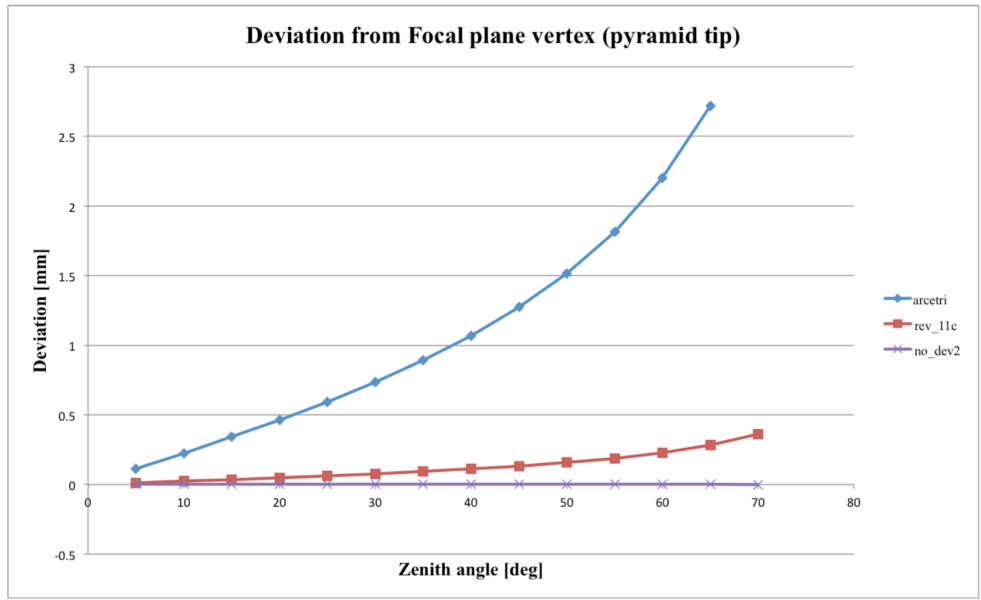

**Fig. 13.** Beam deviation in mm as a function of Zenith angle on the sky. Our as-built triplet ADC is ~90% better than the conventional doublet design at higher airmasses. The zero-deviation triplet design is a further improvement still.

## Appendix

Executing C:\Program Files\ZEMAX\Macros\Z_SCAN_ADC_ANGLE.ZPL.

Altitude in meters is: 2380.0000
Temperature in K is: 288.0000
Pressure in millibars is: 765.0000
Humidity fraction is: 0.2500

First column is zenith angle
Second column is clocking angle of ADC1
Third column is clocking angle of ADC2
Standing at the tertiary looking towards the W-unit,
a positive angle is clockwise.

Wavelength band is: 0.6000-1.0000 microns

1.0000 0.2832 -0.2832
2.0000 0.5665 -0.5665
3.0000 0.8502 -0.8502
4.0000 1.1344 -1.1344
5.0000 1.4193 -1.4193
6.0000 1.7051 -1.7051
7.0000 1.9919 -1.9919
8.0000 2.2803 -2.2803
9.0000 2.5700 -2.5700
10.0000 2.8612 -2.8612
11.0000 3.1545 -3.1545
12.0000 3.4498 -3.4498
13.0000 3.7473 -3.7473
14.0000 4.0474 -4.0474
15.0000 4.3502 -4.3502
16.0000 4.6559 -4.6559
17.0000 4.9649 -4.9649
18.0000 5.2773 -5.2773
19.0000 5.5934 -5.5934
20.0000 5.9135 -5.9135
21.0000 6.2378 -6.2378
22.0000 6.5668 -6.5668
23.0000 6.9006 -6.9006
24.0000 7.2396 -7.2396
25.0000 7.5841 -7.5841
26.0000 7.9346 -7.9346
27.0000 8.2913 -8.2913
28.0000 8.6548 -8.6548
29.0000 9.0255 -9.0255
30.0000 9.4037 -9.4037
31.0000 9.7900 -9.7900
32.0000 10.1849 -10.1849
33.0000 10.5890 -10.5890
34.0000 11.0029 -11.0029

```
35.0000 11.4271 -11.4271
36.0000 11.8626 -11.8626
37.0000 12.3098 -12.3098
38.0000 12.7697 -12.7697
39.0000 13.2432 -13.2432
40.0000 13.7310 -13.7310
41.0000 14.2344 -14.2344
42.0000 14.7543 -14.7543
43.0000 15.2921 -15.2921
44.0000 15.8490 -15.8490
45.0000 16.4264 -16.4264
46.0000 17.0260 -17.0260
47.0000 17.6498 -17.6498
48.0000 18.2992 -18.2992
49.0000 18.9768 -18.9768
50.0000 19.6849 -19.6849
51.0000 20.4263 -20.4263
52.0000 21.2037 -21.2037
53.0000 22.0209 -22.0209
54.0000 22.8812 -22.8812
55.0000 23.7897 -23.7897
56.0000 24.7509 -24.7509
57.0000 25.7708 -25.7708
58.0000 26.8558 -26.8558
59.0000 28.0140 -28.0140
60.0000 29.2543 -29.2543
61.0000 30.5875 -30.5875
62.0000 32.0267 -32.0267
63.0000 33.5874 -33.5874
64.0000 35.2889 -35.2889
65.0000 37.1552 -37.1552
66.0000 39.2167 -39.2167
67.0000 41.5132 -41.5132
68.0000 44.0988 -44.0988
69.0000 47.0470 -47.0470
70.0000 50.4677 -50.4677
DONE.
```